\documentclass[showpacs, showkeys,twocolumn,amsmath,amssymb,prb]{revtex4}

\usepackage{graphicx}
\usepackage{dcolumn}
\usepackage{bm}
\begin{document}
\newcommand{\ud}{d}
\newcommand{\hs}{\hspace*{0.5cm}}


\title{Energy estimators for random series path-integral methods}

\author{Cristian Predescu}
\author{Dubravko Sabo}
\author{J. D. Doll}
\affiliation{Department of Chemistry,  Brown University,  Providence,
Rhode Island 02912}
\author{David L. Freeman}
\affiliation{Department of Chemistry, University of Rhode Island,
Kingston, Rhode Island 02881}
\date{\today}

\begin{abstract} We perform a thorough analysis on the choice of
estimators for random series path integral methods. In particular, we
show that both the thermodynamic (T-method) and the direct (H-method)
energy estimators have finite variances and are straightforward to
implement. It is demonstrated that the agreement between the T-method and
the H-method estimators provides an important consistency check on the
quality of the path integral simulations. We illustrate the behavior of
the various estimators by computing the total, kinetic, and potential
energies of a molecular hydrogen cluster using three different path
integral techniques. Statistical tests are employed to validate the
sampling strategy adopted as well as to measure the performance of the
parallel random number generator utilized in the Monte Carlo simulation.
Some issues raised by previous simulations of the hydrogen cluster are
clarified.
\end{abstract}


\maketitle

\section{Introduction} \label{sec:intro}

Numerical path integral methods have proved to be highly useful tools in
the analysis of finite temperature, many-body quantum
systems.\cite{Nig99}  A central theme in such studies is the conscious
use of dimensionality, both in the reformulation of the original problem
and in the subsequent numerical simulations.

As the scale of the problems under study continues to grow, it becomes
increasingly important that the formal properties of the numerical
methods that are utilized be properly characterized.  Recently, Predescu
and co-workers\cite{Pre02, Pre03a, Pre03} have presented a number of
results concerning the convergence properties of random series-based path
integral techniques.  Important in their own right, these formal
properties have also led to the development of a new class of path
integral methods, the so-called reweighted techniques.\cite{Pre03}
Reweighted approaches accelerate the convergence of ``primitive'' series
methods by including the effects of ``higher-order'' path variables in a
simple, approximate fashion.  Reweighted methods achieve  the convergence
rate of related partial averaging approaches\cite{Dol85}  without
requiring the construction of the Gaussian transform of the underlying
potential energy function.

Previous work on the reweighted method  has focused principally on the
construction of the quantum-mechanical density matrix.\cite{Pre03,
Pre03b}  In the present work, we wish to examine estimators for various
coordinate-diagonal and off-diagonal properties.  While the present
discussion is focused principally on reweighted methods, the results
obtained are broadly applicable to more general random series approaches.

In Section II of the present article, we examine
the thermodynamic (T-method) and direct (H-method) estimators for the
total energy. In order to avoid any confusion with earlier estimators, we mention that in the present article by  T-method and H-method estimators we understand the respective energy estimators introduced by Predescu and Doll in Ref.~\onlinecite{Pre02}.  Thus, the T-method estimator we employ does not have the  variance  difficulties associated with  the Barker estimator for large numbers of path variables.\cite{Her82} As the low-temperature simulation presented in the second part of the article demonstrates, the present T-method estimator does not exhibit any of the difficulties sometimes associated with the virial estimator for low-temperature systems or for strongly correlated Monte Carlo sampling techniques.\cite{Gia88, Fer95, Cao89, Kol96, Jan97} The T-method estimator is closely related and similar in form to the centroid virial estimator.\cite{Cep95, Gla02} We expect the two estimators to have  similar behavior with the nature of the quantum system, the temperature, and the Monte Carlo sampling method. However, an important difference between the two estimators is the fact that the T-method estimator is a veritable thermodynamic estimator, in the sense that it is obtained by temperature differentiation of the quantum partition function. This observation is important because the temperature differentiation can be  implemented numerically by a finite-difference scheme and, in principle, may lead to  numerically stable algorithms that do not require derivatives of the potential. For large dimensional systems or systems described by complicated potentials, we expect such algorithms to be significantly faster than those based on explicit analytical formulas.  The relative merits of such algorithms will be examined in future work. 

 In Section III, we examine the application
of the reweighted methods to a model problem, that of simulating the
thermodynamic properties of the (H$_2)_{22} $ molecular cluster.
In Section IV, we summarize our present findings and clarify a
number of issues raised in previous studies of this molecular hydrogen
system.\cite{Cha98, Dol99}

\section{Energy estimators} \label{sec:estimators} In this section, we
consider a one-dimensional quantum canonical system characterized by
inverse temperature $\beta = 1/ (k_{B} T)$ and set forward the task of
computing its average energy by Monte Carlo integration methods developed
around several reweighted techniques.\cite{Pre03, Pre03b} The physical
system is made up of a particle of mass $m_0$ moving in the potential
$V(x)$. We discuss the numerical implementation and the computational
merits of both the T-method and H-method estimators. Any time the
multidimensional extension is not obvious, we present the explicit
formulas of the respective estimators.

We begin by presenting the general form of  the  path integral methods we
employ in this paper. We remind the reader that in terms of a standard
Brownian motion $\left\{B_u, u \geq 0\right\}$, the Feynman-Ka\c{c}
formula has the expression\cite{Sim79}
\begin{eqnarray}
\label{eq:1a}&& \nonumber
\rho(x,x';\beta) = P\left[\sigma B_1 = x' | \sigma B_0 = x \right] \\
&&\times \mathbb{E}\left[e^{-\beta \int_0^1 V(\sigma B_u)\ud u } | \sigma
B_1 = x' , \sigma B_0 = x \right],
\end{eqnarray} where $\sigma = (\hbar^2\beta/m_0)^{1/2}$. In this paper, we shall use the symbol $\mathbb{E}$ to denote the expected value (average value) of a certain random variable against the underlying probability measure of the Brownian motion $B_u$.   It is
straightforward to see that the first factor of the product in
Eq.~(\ref{eq:1a}) (which represents the conditional probability density
that the rescaled Brownian motion $\sigma B_u$ reaches the point $x'$
provided that it starts at the point $x$) is the density matrix of a free
particle of mass $m_0$
\[ P\left[\sigma B_1 = x' | \sigma B_0 = x \right] =
\rho_{fp}(x,x';\beta).
\] Moreover, rather than using the conditional expectation appearing in
the second factor of Eq.~(\ref{eq:1a}), one usually employs a stochastic
process $\{ B_u^0; 0 \leq u \leq 1\}$, called a standard Brownian
bridge,\cite{Sim79, Dur96} which is defined as a standard Brownian motion
conditioned on the end points such that $B_0^0 = 0$ and $B_1^0 = 0$. In
terms of the newly defined process, the Feynman-Ka\c{c} formula reads
\[
\frac{\rho(x,x';\beta)}{\rho_{fp}(x,x';\beta)}= \mathbb{E}
\exp\left\{-\beta \int_0^1 V[x_r(u)+\sigma B_u^0]\ud u\right\},
\] where $x_r(u) = x + (x'-x)u$ is a straight line connecting the points
$x$ and $x'$ and is called the reference path.

   As discussed in Ref.~\onlinecite{Pre02}, one of the most general
constructions of the standard Brownian bridge is given by the Ito-Nisio
theorem.\cite{Kwa92}  Let $\{\lambda_k(\tau)\}_{k \geq 1}$ be a system of
functions on the interval $[0,1]$, which together with the constant
function $\lambda_0(\tau)=1$, make up an orthonormal basis in $L^2[0,1]$.
    Let
\[ \Lambda_k(t)=\int_0^t \lambda_k(u)\ud u.\] If $\Omega$ is the space of
infinite sequences $\bar{a}\equiv(a_1,a_2,\ldots)$ and
\begin{equation}
\label{eq:1}
\ud P[\bar{a}]=\prod_{k=1}^{\infty}\ud \mu(a_k)
\end{equation}
    is the probability measure on $\Omega$ such that the coordinate maps
$\bar{a}\rightarrow a_k$ are independent identically distributed (i.i.d.)
variables with distribution probability
\begin{equation}
\label{eq:2}
\ud \mu(a_i)= \frac{1}{\sqrt{2\pi}} e^{-a_i^2/2}\,\ud a_i,
\end{equation} then
\begin{equation}
\label{eq:2a} B_u^0(\bar{a})\stackrel{d}{=}
\sum_{k=1}^{\infty}a_k\Lambda_{k}(u),\; 0\leq u\leq1 ;
\end{equation} i.e., the right-hand side random series is equal in
distribution to a standard Brownian bridge.  The
notation~$B_u^0(\bar{a})$ in~(\ref{eq:2a}) is then appropriate and allows
us to interpret the Brownian bridge as a collection of random functions
of argument~$\bar{a}$, indexed by~$u$.

Using the Ito-Nisio representation of the Brownian bridge,  the
Feynman-Ka\c{c} formula takes the form
\begin{eqnarray}
\label{eq:3}
    \frac{\rho(x, x' ;\beta)}{\rho_{fp}(x, x'  ;\beta)}&=&\int_{\Omega}\ud
P[\bar{a}]\nonumber  \exp\bigg\{-\beta
\int_{0}^{1}\! \!  V\Big[x_r(u) \\& +& \sigma \sum_{k=1}^{\infty}a_k
\Lambda_k(u) \Big]\ud u\bigg\}.
\end{eqnarray} For a multidimensional system, the Feynman-Ka\c{c} formula
is obtained by employing an independent random series for each additional
degree of freedom.

A reweighted method constructed from the random series $\sum_{k=1}^\infty
a_k \Lambda_k(u)$ is any sequence of approximations to the density matrix
of the form\cite{Pre03}
\begin{eqnarray}
\label{eq:5}&&
\frac{\rho^{\text{RW}}_n(x, x' ;\beta)}{\rho_{fp}(x, x'
;\beta)}=\int_{\mathbb{R}}\ud \mu(a_1)\ldots \int_{\mathbb{R}}\ud
\mu(a_{qn+p})\nonumber  \\&& \times \exp\bigg\{-\beta \; \int_{0}^{1}\! \!
V\Big[x_r(u)+ \sigma \sum_{k=1}^{qn+p}a_k
\tilde{\Lambda}_{n,k}(u)\Big]\ud u\bigg\},\qquad
\end{eqnarray} where $q$ and $p$ are some fixed integers, where
\begin{equation}
\label{eq:6}
\tilde{\Lambda}_{n,k}(u)= \Lambda_k(u) \quad \text{if} \ 1\leq k \leq n,
\end{equation}  and where
\begin{equation}
\label{eq:7}
\sum_{k=n+1}^{qn+p}\tilde{\Lambda}_{n,k}(u)^2=\sum_{k=n+1}^{\infty}
\Lambda_{k}(u)^2.
\end{equation} 
In Eq.~(\ref{eq:5}), $n$ indexes the sequence of reweighted approximations $\rho^{\text{RW}}_n(x, x' ;\beta)$, sequence that converges to the density matrix $\rho(x, x' ;\beta)$ in the limit $n\to \infty$. Remark that the approximation of index $n$ actually utilizes $qn+p$ variables for path parameterization. In the construction of a certain path, the first $n$ functions $\tilde{\Lambda}_{n,k}(u)$ coincide with the ones for the corresponding series representation, as shown by Eq.~(\ref{eq:6}).  A number of $(q-1)n+p$ additional functions are constructed so that to maximize the order of convergence of the reweighted approximation. Notice that if the resulting approximation has a convergence of order $\alpha$ as measured against $n$, then it has the same order of convergence when measured against the total number of variables $qn+p$, though the convergence constant is $q^\alpha$ times larger. This explains why the number of additional functions is chosen to scale linearly with $n$. For additional information, the reader is advised to consult Ref.~\onlinecite{Pre03}. 

  It is convenient to introduce the additional quantities
$X_n(x,x',\bar{a};\beta)$ and $X_\infty(x,x',\bar{a};\beta)$, which are
defined by the expressions
\begin{eqnarray}
\label{8}&& \nonumber X_n(x,x',\bar{a};\beta)=\rho_{fp}(x,x';\beta)\\&&
\times \exp\bigg\{-\beta \; \int_{0}^{1}\! \! V\Big[x_r(u)+ \sigma
\sum_{k=1}^{qn+p}a_k
\tilde{\Lambda}_{n,k}(u)\Big]\ud u\bigg\}\qquad
\end{eqnarray} and
\begin{eqnarray}
\label{9}&& \nonumber
X_\infty(x,x',\bar{a};\beta)=\rho_{fp}(x,x';\beta)\\&&
\times \exp\bigg\{-\beta \; \int_{0}^{1}\! \! V\Big[x_r(u)+ \sigma
\sum_{k=1}^{\infty}a_k
\Lambda_{k}(u)\Big]\ud u\bigg\},\qquad
\end{eqnarray} respectively. With the new notation, Eq.~(\ref{eq:5})
becomes
\begin{eqnarray}
\label{eq:10}&&
\rho^{\text{RW}}_n(x, x' ;\beta)=\int_{\Omega}\ud
P[\bar{a}]X_n(x,x',\bar{a};\beta),
\end{eqnarray} while the Feynman-Ka\c{c} formula reads
\begin{eqnarray}
\label{eq:11}&&
\rho(x, x' ;\beta)=\int_{\Omega}\ud
P[\bar{a}]X_\infty(x,x',\bar{a};\beta).
\end{eqnarray}

   The analytical expressions of the functions
$\tilde{\Lambda}_{n,k}(u)$ depend on the nature of the reweighted
techniques and are generally  chosen to maximize the asymptotic
convergence of the  respective reweighted techniques.\cite{Pre03}  To a
large extent,  the specific form of these functions is not important for
the present development,  but the reader is advised to consult
Refs.~\onlinecite{Pre03} and \onlinecite{Pre03b} for quadrature
techniques and  additional clarifications.

The remainder of the present section is split into two parts. First, we
discuss the problem of computing the ensemble averages of operators
diagonal in coordinate representation. In particular, this resolves the
problem of computing the average potential energy. Second, we consider
the problem of evaluating the total energies (hence, also the kinetic
energies) by means of the T-method and H-method estimators.

\subsection{Operators diagonal in the coordinate representation}

By definition, the ensemble average of an operator $\hat{O}$ diagonal in
the coordinate representation is
\begin{equation}
\label{eq:12}
\left\langle O \right\rangle_{\beta}=\frac{\int_{\mathbb{R}}
\rho(x;\beta)O(x)\ud x}{\int_{\mathbb{R}} \rho(x;\beta)\ud x}.
\end{equation} The quantity $\rho(x;\beta) = \rho(x,x;\beta)$ is the
diagonal density matrix. By convention, we drop the second variable of
the pair $(x,x')$ any time $x = x'$. For instance, we use $X_n(x,
\bar{a};\beta)$ instead of $X_n(x, x,\bar{a};\beta)$. By means of
Eq.~(\ref{eq:11}), the average above can be recast as
\begin{equation}
\label{eq:13}
\left\langle O \right\rangle_{\beta}=\frac{\int_{\mathbb{R}}\ud x
\int_{\Omega}\ud
P[\bar{a}]X_\infty(x,\bar{a};\beta)O(x)}{\int_{\mathbb{R}} \ud x
\int_{\Omega}\ud P[\bar{a}]X_\infty(x,\bar{a};\beta)}.
\end{equation} This average can be recovered as the limit $n \to \infty$
of the sequence
\begin{equation}
\label{eq:14}
\left\langle O
\right\rangle_{\beta,n}^{\text{pt}}=\frac{\int_{\mathbb{R}}\ud x
\int_{\Omega}\ud P[\bar{a}]X_n(x,\bar{a};\beta)O(x)}{\int_{\mathbb{R}}
\ud x \int_{\Omega}\ud P[\bar{a}]X_n(x,\bar{a};\beta)},
\end{equation} the terms of which are to be evaluated by Monte Carlo
integration. The estimating function $O(x)$ appearing in the above
formula is called the point estimating function of the operator
$\hat{O}$.

An alternative to the point estimating function is the so-called path
estimating function, the derivation of which is presented shortly. As
demonstrated in Appendix A, the function $O(x)$ appearing in
Eq.~(\ref{eq:13}) can be replaced by $O[x+ \sigma B_u^0(\bar{a})]$,
without changing the value of the average $\left\langle O
\right\rangle_{\beta}$. That is, the equality
\[
\left\langle O \right\rangle_{\beta}=\frac{\int_{\mathbb{R}}\ud x
\int_{\Omega}\ud P[\bar{a}]X_\infty(x,\bar{a};\beta)O[x+ \sigma
B_u^0(\bar{a})]}{\int_{\mathbb{R}} \ud x \int_{\Omega}\ud
P[\bar{a}]X_\infty(x,\bar{a};\beta)}
\] is valid for all $0 \leq u \leq 1$. Averaging over the variable $u$,
one obtains
\begin{equation}
\label{eq:15}
\left\langle O \right\rangle_{\beta}=\frac{\int_{\mathbb{R}}\ud x
\int_{\Omega}\ud P[\bar{a}]X_\infty(x,\bar{a};\beta)\int_0^1 O[x+ \sigma
B_u^0(\bar{a})] \ud u}{\int_{\mathbb{R}} \ud x \int_{\Omega}\ud
P[\bar{a}]X_\infty(x,\bar{a};\beta)}.
\end{equation} Eq.~(\ref{eq:15}) shows that the ensemble average of the
operator $\hat{O}$ can also be recovered as the limit $n \to \infty$ of
the sequence
\begin{equation}
\label{eq:16}
\left\langle O
\right\rangle_{\beta,n}^{\text{pth}}=\frac{\int_{\mathbb{R}}\ud x
\int_{\Omega}\ud P[\bar{a}]X_n(x,\bar{a};\beta)\int_0^1 O[x+ \sigma
\tilde{B}_{u,n}^0(\bar{a})] \ud u}{\int_{\mathbb{R}} \ud x
\int_{\Omega}\ud P[\bar{a}]X_n(x,\bar{a};\beta)},
\end{equation} where we have set
\[
\tilde{B}_{u,n}^0(\bar{a})= \sum_{k=1}^{qn+p}a_k
\tilde{\Lambda}_{n,k}(u)
\] for convenience of notation.

In the remainder of the present subsection, we discuss the relative
merits of the point and path estimators. We first consider which of
$\left\langle O \right\rangle_{\beta,n}^{\text{pt}}$ and $\left\langle O
\right\rangle_{\beta,n}^{\text{pth}}$ is closer to $\left\langle O
\right\rangle_\beta$ for a given $n$ assuming the averages given in Eqs.
(\ref{eq:14}) and (\ref{eq:16}) are computed exactly. Let us notice that
Eq.~(\ref{eq:14}) can be put in the form
\[
\left\langle O
\right\rangle_{\beta,n}^{\text{pt}}=\frac{\int_{\mathbb{R}}\ud x
\rho_n^{\text{RW}}(x;\beta)O(x)}{\int_{\mathbb{R}} \ud x
\rho_n^{\text{RW}}(x;\beta)}.
\] The probability distribution
\begin{equation}
\label{eq:17}
\frac{\rho_n^{\text{RW}}(x;\beta)\ud x}{\int_{\mathbb{R}}
\rho_n^{\text{RW}}(x;\beta) \ud x }
\end{equation}
   represents the marginal distribution of the  variable $x$ regarded as a
random variable on the space $\mathbb{R} \times \Omega$, which is endowed
with the probability measure
\begin{equation}
\label{eq:18}
\frac{X_n(x,\bar{a};\beta) \ud x \ \ud P[\bar{a}]}{\int_{\mathbb{R}} \ud
x \int_{\Omega}\ud P[\bar{a}]X_n(x,\bar{a};\beta)}.
\end{equation} The reweighted techniques are designed so that the
distribution given by Eq.~(\ref{eq:17}) is as close as possible to the
quantum statistical one, which is given by the expression
\[\frac{\rho(x;\beta)\ud x}{\int_{\mathbb{R}}\rho(x;\beta)\ud x }.\]  In
designing the reweighted techniques, one seeks to optimize the rate of
convergence  of the sequence $\rho_n^{\text{RW}}(x,x';\beta) \to
\rho(x,x';\beta)$ for all $x$ and $x'$.\cite{Pre03}

For arbitrary $u$, the marginal distribution of $x+ \sigma
B_{u,n}^0(\bar{a})$ is usually different from the one given by
Eq.~(\ref{eq:17}) and is not optimized. With few notable exceptions to be
analyzed below, the points $x+ \sigma B_{u,n}^0(\bar{a})$ for different
$u$ are not equivalent, and their probability distribution may differ
significantly from the quantum statistical one. (However, as shown in
Appendix~A, they become equivalent in the limit $n \to \infty$.)
Therefore, especially for those reweighted techniques having fast
asymptotic convergence, we expect the point estimator to be more rapidly
convergent with $n$ than the path estimator.

An additional issue appearing in Monte Carlo computations is the variance
of the two estimating functions $O(x)$ and $\int_0^1 O[x+ \sigma
\tilde{B}_{u,n}^0(\bar{a})]\ud u$. In the limit $n \to \infty$, the variance of
the point estimating function converges to
\begin{eqnarray*}&&
\frac{\int_{\mathbb{R}}\ud x \int_{\Omega}\ud
P[\bar{a}]X_\infty(x,\bar{a};\beta)O(x)^2}{\int_{\mathbb{R}} \ud x
\int_{\Omega}\ud P[\bar{a}]X_\infty(x,\bar{a};\beta)}-\left\langle O
\right\rangle_{\beta}^2 \\ && =
\frac{\int_{\mathbb{R}}\ud x \int_{\Omega}\ud
P[\bar{a}]X_\infty(x,\bar{a};\beta)\int_0^1 O[x+ \sigma
B_{u}^0(\bar{a})]^2 \ud u}{\int_{\mathbb{R}} \ud x \int_{\Omega}\ud
P[\bar{a}]X_\infty(x,\bar{a};\beta)} \\ && \ \ \ \ \ \ \ \ \ \
-\left\langle O \right\rangle_{\beta}^2,
\end{eqnarray*} while the variance of the path estimating function
converges to
\[\frac{\int_{\mathbb{R}}\ud x \int_{\Omega}\ud
P[\bar{a}]X_\infty(x,\bar{a};\beta)\left\{\int_0^1 O[x+ \sigma
{B}_{u}^0(\bar{a})] \ud u\right\}^2}{\int_{\mathbb{R}} \ud x
\int_{\Omega}\ud P[\bar{a}]X_\infty(x,\bar{a};\beta)} -\left\langle O
\right\rangle_{\beta}^2.
\] The Cauchy-Schwartz inequality implies
\[
\left\{\int_0^1 O[x+ \sigma {B}_{u}^0(\bar{a})] \ud u\right\}^2 \leq
\int_0^1 O[x+ \sigma {B}_{u}^0(\bar{a})]^2 \ud u
\] and shows that the variance of the path estimating function is always
smaller than that of the point estimating function. The actual decrease
in the variance is not always significant because the points $x+ \sigma
{B}_{u}^0(\bar{a})$ for different $u$ are strongly correlated. Depending
on the nature of the function $O(x)$, the variance decrease may not
compensate the effort required to compute the average $\int_0^1 O[x+
\sigma \tilde{B}_{u,n}^0(\bar{a})] \ud u$. However, if the function
$O(x)$ is the potential $V(x)$, then the smaller variance of the path
estimator is a desirable feature because the  path average $\int_0^1 V[x+
\sigma \tilde{B}_{u,n}^0(\bar{a})] \ud u$, which also enters the
expression of $X_n(x,\bar{a};\beta)$, is computed anyway.

To summarize the findings of the present subsection, the point estimator
provides a more accurate value but has a larger variance than the path
estimator.  We next ask if there are any methods for which one may construct
an estimator providing the same values as the point estimator but having
the variance of the path estimator. More precisely, we seek methods for
which there is a division
$0= u_0 \leq u_1 \leq \ldots \leq u_{q_n} \leq u_{q_n+1} = 1$ such that
the mesh $\max_{0 \leq i \leq q_n} |u_{i+1} - u_i|$ converges to zero as
$n \to \infty$ and such that the points
$\left\{x+ \sigma \tilde{B}_{u_i,n}^0(\bar{a}); 0 \leq i \leq
q_n+1\right\}$ have the same  marginal distribution as $x$. For such
methods, the expected value  of the estimating function
\begin{equation}
\label{eq:19}
\sum_{i = 0}^{q_n} O[x+ \sigma
\tilde{B}_{u_i,n}^0(\bar{a})](u_{i+1}-u_{i})
\end{equation} under the probability distribution given by
Eq.~(\ref{eq:18}) is an estimator satisfying the criteria outlined in
this paragraph.

There are two methods we employ in the present paper for which such an
estimator exists. The first one, is the trapezoidal Trotter discrete path
integral method (TT-DPI) obtained by the Trotter composition
\begin{eqnarray}
\label{eq:20}
\rho_n^{\text{TT}}(x,x';\beta)=\int_{\mathbb{R}}\ud x_1 \ldots
\int_{\mathbb{R}}\ud x_n\;
\rho_0\left(x,x_1;\frac{\beta}{n+1}\right)\nonumber \\ \ldots
\rho_0\left(x_n,x';\frac{\beta}{n+1}\right)\qquad
\end{eqnarray} of the short-time approximation
\[
\rho_0^{\text{TT}}(x,x';\beta) = \rho_{fp}(x, x';\beta) \exp\left[-\beta
\frac{V(x)+V(x')}{2} \right].
\]
   It has been shown\cite{Pre02b} that for $n=2^k-1$, the TT-DPI method
admits the following   implementation
\begin{widetext}
\begin{eqnarray}
\label{eq:21}\nonumber
\frac{\rho_n^{\text{TT}}(x, x' ;\beta)}{\rho_{fp}(x, x'
;\beta)}=\int_{\mathbb{R}}\ud a_{1,1}\ldots \int_{\mathbb{R}}\ud
a_{k,2^{k-1}}  \left( 2\pi \right)^{-n/2}
\exp\left({-\frac{1}{2}\sum_{l=1}^k\sum_{i=1}^{2^{l-1}} a_{l,i}^2}\right)
\\ \times \exp\left\{-{\beta}\sum_{i=0}^{2^k}\omega_i
V\left[x_r(u_i)+\sigma \sum_{l=1}^{k}F_{l,[2^{l-1}
u_i]+1}(u_i)a_{l,[2^{l-1} u_i]+1}\right]\right\},
\end{eqnarray}
\end{widetext} where $u_i= 2^{-k} i $ for $0\leq i \leq 2^k$ and
\[
\omega_i =\left\{ \begin{array}{l l}2^{-(k+1)},& \text{if}\ i\in \{0,
2^k\},\\ 2^{-k}, & \text{if}\ 1\leq i \leq 2^k-1.
   \end{array} \right.
\] The functions $F_{l,k}(u)$ are the so-called Schauder
functions,\cite{McK69} the definitions of which are presented in the
cited references. We leave it for the reader to use Eq.~(\ref{eq:20})
  and show that if $x = x'$, then all the points $ x+ \sigma
\tilde{B}_{u_i,n}^0(\bar{a})$ have identical marginal distribution given
by the formula
\[\frac{\rho_n^{\text{TT}}(x;\beta)\ud
x}{\int_{\mathbb{R}}\rho_n^{\text{TT}}(x;\beta)\ud x }.\] In this case,
the point and the path estimators produce identical results for the
ensemble average of a diagonal operator $\hat{O}$
\[\frac{\int_\mathbb{R}\rho_n^{\text{TT}}(x;\beta) O(x) \ud
x}{\int_{\mathbb{R}}\rho_n^{\text{TT}}(x;\beta)\ud x }.\] At least for
the ensemble average of the potential energy, one should always use the
path estimator, which has smaller variance.

A second method for which there is an estimator giving the same values as
the point estimator but having (asymptotically, as $n \to \infty$) the
variance of the path estimator is the so-called L\'evy-Ciesielski
reweighted technique (RW-LCPI) defined by the formula\cite{Pre03}
\begin{widetext}
\begin{eqnarray}
\label{eq:22} \nonumber
\frac{\rho_n^{\text{LC}}(x, x' ;\beta)}{\rho_{fp}(x, x'
;\beta)}=&&\int_{\mathbb{R}}\ud a_{1,1}\ldots \int_{\mathbb{R}}\ud
a_{k+2,2^{k+1}}  \left( 2\pi \right)^{-(4n+3)/2}
\exp\left({-\frac{1}{2}\sum_{l=1}^{k+2}\sum_{j=1}^{2^{l-1}}
a_{l,j}^2}\right)
\\&& \times \exp\left\{-\beta \int_0^1 V\left[x_r(u)+\sigma
\sum_{l=1}^{k+2} a_{l,[2^{l-1} u]+1} \;\tilde{F}^{(n)}_{l,[2^{l-1}
u]+1}(u)\right]\ud u\right\},
\end{eqnarray}
\end{widetext} where $[2^{l-1} u]$ is the integer part of $2^{l-1} u$.
   It has been shown that for $n=2^k-1$, the RW-LCPI method can be put in
the Trotter product form\cite{Pre03}
\begin{eqnarray}
\label{eq:23}
\rho_n^{\text{LC}}(x,x';\beta)=\int_{\mathbb{R}}\ud x_1 \ldots
\int_{\mathbb{R}}\ud x_n\;
\rho_0^{\text{LC}}\left(x,x_1;\frac{\beta}{n+1}\right)\nonumber \\ \ldots
\rho_0^{\text{LC}}\left(x_n,x';\frac{\beta}{n+1}\right),\qquad
\end{eqnarray} where
\begin{eqnarray*}\nonumber
\frac{\rho_0^{\text{LC}}(x,x';\beta)}{\rho_{fp}(x,x';\beta)}=\frac{1}{\left(2\pi\right)^{3/2}}\int_{\mathbb{R}}\int_{\mathbb{R}}\int_{\mathbb{R}}
e^{-\left(a_1^2+a_2^2+a_3^2\right)/2}\\\times \exp\bigg\{-\beta \int_0^1
V[x+(x'-x)u+a_1\sigma C_0(u)\\ +a_2 \sigma L_0(u)+ a_3 \sigma R_0(u)] \ud
u\bigg\} \ud a_1 \ud a_2 \ud a_3. \nonumber
\end{eqnarray*} The analytical expressions of the functions
$\tilde{F}^{(n)}_{k,l}(u)$, $L_0(u)$, $R_0(u)$, and $C_0(u)$ can be found
in Refs.~\onlinecite{Pre03} and \onlinecite{Pre03b}.

Again, we leave it for the reader to use Eq.~(\ref{eq:23}) and prove that
if $x'=x$, then all the points
\[ x+\sigma \sum_{l=1}^{k+2} a_{l,[2^{l-1} u_i]+1}
\;\tilde{F}^{(n)}_{l,[2^{l-1} u_i]+1}(u_i)
\] with $u_i = 2^{-k}i$ for $0 \leq i \leq 2^k$ have identical marginal
distributions equal to that of $x$. The estimator
\begin{equation}
\label{eq:24} 2^{-k}\sum_{i = 0}^{2^{k}-1} O\left[ x+\sigma
\sum_{l=1}^{k+2} a_{l,[2^{l-1} u_i]+1} \;\tilde{F}^{(n)}_{l,[2^{l-1}
u_i]+1}(u_i)\right]
\end{equation} produces the same results as the point estimator, but it
has the variance of the path estimator. As far as the evaluation of the
average potential energy is concerned, in order to avoid unnecessary
calls to the potential routine, it is desirable that the points $\{
2^{-k} i; 0 \leq i \leq 2^k\}$ be among the quadrature points utilized
for the computation of the path averages appearing in Eq.~(\ref{eq:22}).
The quadrature technique designed in Ref.~\onlinecite{Pre03b} shares
this property. As opposed to the TT-DPI method, the point and the path
estimators for the RW-LCPI method produce different results.

\subsection{Estimators for the total energy}  In this subsection, we
discuss the implementation of the thermodynamic (T) and the direct (H)
estimators for the total energy.  The T-method estimator is defined  as
the following functional of the diagonal density matrix:
\begin{equation}
\label{eq:25}
\left\langle E \right\rangle^{T}_{\beta} =-\frac{\partial}{\partial \beta}
\ln{\left[ \int_{\mathbb{R}} \rho(x;\beta)\ud x \right]}.
\end{equation} The above formula can be expressed as the statistical
average
\begin{equation}
\label{eq:26}
\left\langle E \right\rangle^{T}_{\beta} =\frac{\int_{\mathbb{R}}\ud x
\int_{\Omega}\ud
P[\bar{a}]X_\infty(x,\bar{a};\beta)E_\infty^T(x,\bar{a};\beta) }
{\int_{\mathbb{R}}\ud x\int_{\Omega}\ud P[\bar{a}]X_\infty(x,
\bar{a};\beta)},
\end{equation} where the T-method estimating function
$E_\infty^T(x,\bar{a};\beta)$  can be shown to be\cite{Pre02}
\begin{eqnarray}
\label{eq:27} \nonumber
   E_\infty^T(x, \bar{a};\beta)=\frac{1}{2\beta}+ \int_{0}^{1}\! \!
V\left[x+ \sigma {B}_{u}^0(\bar{a}) \right]\,\ud u \\ +
\frac{\sigma}{2}\int_{0}^{1}\! \!   V'\left[x+ \sigma {B}_{u}^0(\bar{a})
\right]{B}_{u}^0(\bar{a})\,\ud u
\end{eqnarray} provided that  $e^{-\beta V(x)}$ has (Sobolev) first order
derivatives as a function of $x$. For a $d$-dimensional system, the
expression of the T-method estimating function reads
\begin{widetext}
\begin{eqnarray}
\label{eq:28} \nonumber
   E_\infty^T(x_1,\ldots,x_d, \bar{a}_1, \ldots,
\bar{a}_d;\beta)=\frac{d}{2\beta}+ \int_{0}^{1}\! \! V\left[x_1+ \sigma_1
{B}_{u}^{0,1}(\bar{a}_1),\ldots, x_d+ \sigma_d {B}_{u}^{0,d}(\bar{a}_d)
\right]\ud u \\ + \sum_{i=1}^d\frac{\sigma_i}{2}\int_{0}^{1}\! \!
\left\{\frac{\partial}{\partial x_i}V\left[x_1+ \sigma_1
{B}_{u}^{0,1}(\bar{a}_1),\ldots, x_d+ \sigma_d {B}_{u}^{0,d}(\bar{a}_d)
\right]\right\}{B}_{u}^{0,i}(\bar{a}_i)\, \ud u.
\end{eqnarray}
\end{widetext}

The ensemble average energy can be obtained as the limit $n \to \infty$
of the sequence
\begin{equation}
\label{eq:29}
\left\langle E \right\rangle^{T}_{\beta,n} =\frac{\int_{\mathbb{R}}\ud x
\int_{\Omega}\ud P[\bar{a}]X_n(x,\bar{a};\beta)E_n^T(x,\bar{a};\beta) }
{\int_{\mathbb{R}}\ud x\int_{\Omega}\ud P[\bar{a}]X_n(x, \bar{a};\beta)},
\end{equation} where
\begin{eqnarray}
\label{eq:30} \nonumber
   E_n^T(x, \bar{a};\beta)=\frac{1}{2\beta}+ \int_{0}^{1}\! \! V\left[x+
\sigma \tilde{B}_{u,n}^0(\bar{a}) \right]\,\ud u \\ +
\frac{\sigma}{2}\int_{0}^{1}\! \!   V'\left[x+ \sigma
\tilde{B}_{u,n}^0(\bar{a}) \right] \tilde{B}_{u,n}^0(\bar{a})\,\ud u.
\end{eqnarray} The finite-dimensional integral appearing in
Eq.~(\ref{eq:29}) can be evaluated by Monte Carlo integration. In the limit $n \to
\infty$, the variance of the estimator is finite because the square of
$E_\infty^T(x,\bar{a};\beta)$ given by Eq.~(\ref{eq:20}) is a well
defined function, the average value of which is finite for smooth enough
potentials.

A second energy estimator we employ in the present paper is the H-method
estimator. This direct estimator is defined by the equation
\begin{equation}
\label{eq:31}
\left\langle E \right\rangle^{H}_{\beta} =\frac{ \int_{\mathbb{R}}
\hat{H}_{x'} \rho(x,x';\beta)\big|_{x'=x} \ud x} { \int_{\mathbb{R}}
\rho(x;\beta) \ud x},
\end{equation} where the Hamiltonian of the system $\hat{H}_{x'}$ is
assumed to act on the density matrix through the variable $x'$. By
explicit computation and some integration by parts, the H-method
estimator can be expressed as the statistical average
\begin{equation}
\label{eq:32}
\left\langle E \right\rangle^{H}_{\beta} =\frac{\int_{\mathbb{R}}\ud x
\int_{\Omega}\ud
P[\bar{a}]X_\infty(x,\bar{a};\beta)E_\infty^H(x,\bar{a};\beta) }
{\int_{\mathbb{R}}\ud x\int_{\Omega}\ud P[\bar{a}]X_\infty(x,
\bar{a};\beta)}
\end{equation} of the estimating function\cite{Pre02}
\begin{eqnarray}
\label{eq:33} \nonumber
   E_\infty^H(x, \bar{a};\beta)=\frac{1}{2\beta}+ V(x)+\frac{\hbar^2
\beta^2}{4 m_0}
\int_{0}^{1}\! \! \int_{0}^{1}\! \!(u-\tau)^2 \\
\times V'[x+\sigma B_u^0(\bar{a})]\, V'[x+\sigma B_\tau^0(\bar{a})]\,\ud
u \,\ud \tau.
\end{eqnarray} The H-estimator is properly defined even  for potentials
that do not have second-order derivatives. For a $d$-dimensional system,
the H-method estimating function reads
\begin{widetext}
\begin{eqnarray}
\label{eq:34}
\nonumber  E_\infty^H(x_1,\ldots,x_d, \bar{a}_1, \ldots,
\bar{a}_d;\beta)&=&\frac{d}{2\beta}+ V\left(x_1,\ldots, x_d \right)  +
\sum_{i = 1}^d \frac{\hbar^2 \beta^2}{4 m_{0,i}}
\int_{0}^{1}\! \! \int_{0}^{1}\! \!(u-\tau)^2 \\  &\times& \left\{
\frac{\partial}{\partial x_i}V\left[x_1+ \sigma_1
{B}_{u}^{0,1}(\bar{a}_1),\ldots, x_d+ \sigma_d {B}_{u}^{0,d}(\bar{a}_d)
\right]\right\} \\&\times& \left\{
\frac{\partial}{\partial x_i}V\left[x_1+ \sigma_1
{B}_{\tau}^{0,1}(\bar{a}_1),\ldots, x_d+ \sigma_d
{B}_{\tau}^{0,d}(\bar{a}_d) \right]\right\}\ud u \, \ud \tau . \nonumber
\end{eqnarray}
\end{widetext}  The reader should notice that the double integral
appearing in Eq.~(\ref{eq:33}) is really a sum of products of one
dimensional integrals. Indeed, one easily computes
\begin{eqnarray}
\label{eq:35} \nonumber && E_\infty^H(x, \bar{a};\beta)=\frac{1}{2\beta}+
V(x)+\frac{\hbar^2 \beta^2}{2 m_0}\\&& \times
\left\{\int_{0}^{1} u^2 V'[x+\sigma B_u^0(\bar{a})]\ud u\right\} \nonumber
\left\{\int_{0}^{1} V'[x+\sigma B_u^0(\bar{a})]\ud u\right\}\\ &&
-\frac{\hbar^2 \beta^2}{2 m_0}\left\{\int_{0}^{1} u V'[x+\sigma
B_u^0(\bar{a})]\ud u\right\}^2.
\end{eqnarray}
  The H-method estimator is the sum of the ``classical'' energy and a
``quantum'' correction term. Equation~(\ref{eq:32}) shows that the total
energy can also be recovered as the limit $n \to \infty$ from the sequence
\begin{equation}
\label{eq:36}
\left\langle E \right\rangle^{H}_{\beta,n} =\frac{\int_{\mathbb{R}}\ud x
\int_{\Omega}\ud P[\bar{a}]X_n(x,\bar{a};\beta)E_n^H(x,\bar{a};\beta) }
{\int_{\mathbb{R}}\ud x\int_{\Omega}\ud P[\bar{a}]X_n(x, \bar{a};\beta)},
\end{equation} where
\begin{eqnarray}
\label{eq:37} \nonumber
   E_n^H(x, \bar{a};\beta)=\frac{1}{2\beta}+ V(x)+\frac{\hbar^2 \beta^2}{4
m_0}
\int_{0}^{1}\! \! \int_{0}^{1}\! \!(u-\tau)^2 \\
\times V'[x+\sigma \tilde{B}_{u,n}^0(\bar{a})]\, V'[x+\sigma
\tilde{B}_{\tau,n}^0(\bar{a})]\,\ud u \,\ud \tau.
\end{eqnarray}
The forms of the T- and the H-method estimators derived here with the
reweighted techniques in mind extend naturally to the TT-DPI method by
means of Eq.~(\ref{eq:21}). One just replaces the one dimensional
integrals appearing in Eqs.~(\ref{eq:30}) and (\ref{eq:37}) by
appropriate trapezoidal quadrature sums. 

For the reweigthed techniques, we anticipate that the kinetic energy estimator entering the H-method estimator provides more accurate results than the kinetic energy estimator entering the T-method estimator. As for the point and the path estimators of diagonal operators,  the derivatives of the density matrix against the spatial coordinates, which measure fluctuations around the preferential points $x$ and $x'$ for which the reweighted density matrices are optimized, are expected to be reproduced in a better way than the temperature derivatives, which involve unoptimized path-averaged fluctuations. However, for sufficiently low temperatures, the variance of the H-method kinetic energy estimator is expected to be larger than the variance of its thermodynamic counterpart. This larger variance is due to the factor $\beta^2$ appearing in Eqs.~(\ref{eq:33}) and (\ref{eq:37}).

There is one special property of the T- and H-method estimators that
proves to be important in  simulations. Let us notice that by virtue of
the Bloch equation
\[
\hat{H}_{x'} \rho(x,x';\beta) = -\frac{\partial}{\partial
\beta}\rho(x,x';\beta),
\] we have the equality
\[
\left\langle E \right\rangle_{\beta}:= \left\langle E
\right\rangle^{H}_{\beta}=\left\langle E \right\rangle^{T}_{\beta}.
\] Here, the symbol $:=$ signifies that the average energy $\left\langle E \right\rangle_{\beta}$ is \emph{defined} to be the common value of the T-method and the H-method energy estimators, provided that these are equal.  However, since $\rho_n^{\text{RW}}(x,x';\beta)$ does not satisfy the
Bloch equation (except for the free particle), in general
\begin{eqnarray*}
\left\langle E \right\rangle^{H}_{\beta,n}= \frac{ \int_{\mathbb{R}}
\hat{H}_{x'} \rho_n^\text{RW}(x,x';\beta)\big|_{x'=x} \ud x} {
\int_{\mathbb{R}} \rho_n^\text{RW}(x;\beta) \ud x}  \\  \neq \left\langle
E \right\rangle^{T}_{\beta,n}=   -\frac{\partial}{\partial \beta}
\ln{\left[ \int_{\mathbb{R}} \rho_n^\text{RW}(x;\beta)\ud x \right]}
\end{eqnarray*} and the T- and H-method estimators produce the same
result only in the limit $n \to \infty$. Given that the two energy
estimators discussed in the present section can be computed
simultaneously without incurring any computational penalty, we recommend
that the agreement between the T- and the H-method estimators  be used as
a verification tool in actual simulations in order to check the
convergence of various path integral methods. However, we emphasize that the agreement between the T- and the H-method estimators is not a sufficient convergence criterion and in practice, the convergence of different ensemble averages with the number of path variables should also be monitored.

As Eqs.~(\ref{eq:30}) and (\ref{eq:37}) show, the path and the point
estimating functions for the potential energy enter naturally the
expressions of the T- and H-method estimating functions, respectively.
For the purpose of using the agreement between the two energy estimators
as a verification tool for convergence, one should not replace the path
estimating function for the potential energy in the expression  of the
T-method estimator with the point estimating function, even if this may
improve the estimated energy.   For special cases, as for instance the
TT-DPI and RW-LCPI methods discussed in the previous subsection, one may
replace the point estimating function for the potential energy appearing
in the expression of the H-method estimator  with other estimating
functions that produce the same value but have smaller variance.  In this
paper, we replace the point estimating function with the path estimating
function for the TT-DPI method and with the estimating function given by
Eq.~(\ref{eq:24}) for the RW-LCPI method, respectively.

\section{A numerical example}

We have tested the relative merits of the T- and H-method energy
estimators on a cluster of 22 hydrogen molecules at a temperature of 6 K,
using three different path integral methods. Two of these methods, the
trapezoidal Trotter discrete path integral method and a L\'evy-Ciesielski
reweighted technique, have been already presented in the preceding
section. The third method is a Wiener-Fourier reweighted (RW-WFPI)
technique introduced in Ref.~\onlinecite{Pre03}. The numerical
implementation of the methods has been extensively discussed in
Ref.~\onlinecite{Pre03b} by some of us and are not reviewed here.

   The physical system we study has been recently examined by Chakravarty,
Gordillo, and Ceperley\cite{Cha98} as well as by Doll and
Freeman\cite{Dol99} in their comparison of Fourier and
discrete path integral Monte Carlo methods. The total potential energy of
the
$(\text{H}_2)_{22}$ cluster is given by
\begin{equation}
\label{2.1} V_{tot} = \sum_{i<j}^{N} V_{LJ}(r_{ij})+\sum_{i=1}^{N}
V_c(\mathbf{r_i}),
\end{equation} where $V_{LJ}(r_{ij})$ is the pair interaction of
Lennard-Jones potential
\begin{equation}
\label{2.2} V_{LJ}(r_{ij}) = 4\epsilon_{LJ}\left
        [\left( \frac{\sigma_{LJ}}{r_{ij}}\right)^{12}
       -\left( \frac{\sigma_{LJ}}{r_{ij}}\right)^{6}\right]
\end{equation} and V$_{c}(\mathbf{r_i})$ is the constraining potential
\begin{equation}
\label{2.3}
V_c(\mathbf{r_i})=\epsilon_{LJ}\left(\frac{|\mathbf{r_i}-\mathbf{R_{cm}}|}{R_c}\right)^{20}.
\end{equation}
   The values of the Lennard-Jones parameters
$\sigma_{LJ}$ and $\epsilon_{LJ}$ used are 2.96 {\AA} and 34.2 K,
respectively. \cite{Cha98}
$\mathbf{R_{cm}}$ is the coordinate  of the center of mass of the
cluster  and is given by
\begin{equation}
\label{2.4}
\mathbf{R_{cm}}=\frac{1}{N}\sum_{i=1}^N \mathbf{r_i}.
\end{equation} Finally, $R_c=4\sigma_{LJ}$ is the constraining radius.
The role of the constraining potential $V_c(\mathbf{r_i})$ is to prevent
molecules from permanently leaving the cluster since the cluster in
vacuum at any finite temperature is metastable with respect to
evaporation.

At the temperature of $6$~K and at the small densities employed in our computation, the molecules of hydrogen can be described by spherical rotational wave functions, because the majority of the molecules are in the $J=0$ state. To a good approximation, the molecules can be regarded as spherical bosons interacting through isotropic pair potentials. However, a thorough study of parahydrogen clusters has showed that quantum exchange of molecules is small at temperatures greater than $2$~K and that the  hydrogen molecules can be safely treated as distinguishable  particles.\cite{Sin91}

The optimal choice of the parameter $R_c$ for the constraining potential
has been discussed in recent work.\cite{Nei00}  If $R_c$ is taken to be
too small, the properties of the system become sensitive to its choice,
whereas large values of $R_c$ can result in problems attaining an ergodic
simulation.  To facilitate comparisons, in the current work, $R_c$ has
been chosen to be identical to that used in Ref.~\onlinecite{Cha98}.
While this choice of constraining potential can induce ergodicity
problems in calculations of fluctuation quantities like the heat
capacity, we provide evidence below that the simulations in the current
work are ergodic.

The three path integral methods we have employed utilize different
numbers of path variables for a given index $n$. For instance, the TT-DPI
$n$-th order approximation to the density matrix
$\rho_n^{\text{TT}}(x,x';\beta)$ utilizes $n$ path variables for each
physical dimension, whereas $\rho_n^{\text{LC}}(x,x';\beta)$ and
$\rho_n^{\text{WF}}(x,x';\beta)$ utilize $4n+3$ and $4n$ path variables,
respectively. To ensure fair comparison with respect to the number of
path variables employed, we have tabled the total number of variables
$n_v$ used for each physical dimension and not the index $n$.

\subsection{Sampling strategy} We have discussed in Section~II that the
evaluation of the ensemble average of any observable eventually reduces
to the evaluation of the average of a certain estimating function against
the probability distribution
\begin{equation}
\label{eq:3.1}
\frac{X_n(x,\bar{a};\beta) \ud x \, \ud P[\bar{a}]}{\int_{\mathbb{R}} \ud
x \int_{\Omega}\ud P[\bar{a}]X_n(x,\bar{a};\beta)}
\end{equation} or its multidimensional counterpart. This probability
distribution can be sampled with the help of the Metropolis algorithm,
which comprises the following steps.\cite{Met53, Kal86} One initializes
the imaginary-time paths in some fashion. Then, one attempts a trial move
of the paths, which may involve changing several coordinates at a time.
The displacement of the new paths is usually chosen to be relative to the
old paths. To ensure ergodicity, one makes sure that all variables of the
system are eventually moved in a cyclic or a random fashion. The proposed
path is then accepted or rejected with a certain probability. The average
value of the quantity of interest is computed by averaging the values of
the corresponding estimating function evaluated at the current paths.

To establish some notation necessary for our discussion, for each vector
$\mathbf{r}_i= ( x_i, y_i, z_i )$ denoting the physical coordinates of
the particle $i$, we let $\mathbf{\bar{a}}_i = \{\mathbf{a}_{i,1},
\ldots, \mathbf{a}_{i,n_v}\}$ be the collection of path variables
associated with the respective particle. Each
\[\mathbf{a}_{i,k}=\left(a^{x}_{i,k}, a^{y}_{i,k}, a^{z}_{i,k}\right)\]
   is itself a three-dimensional vector whose components denote the  $k$-th
parameter of particle $i$ for the $x$, $y$, and $z$ coordinates,
respectively. Going back to the description of the Metropolis algorithm,
the full imaginary-time path has been initialized by choosing the
physical coordinates $\bf{r}_i$ randomly in a sphere of radius $R_c$
centered about origin. The path variables $\bf{\bar{a}}_i$ have been
initialized with zero.

Except for the Wiener-Fourier method with $n_v = 512$ ($n = 128$), we
update the individual particles one at a time in a cyclic fashion. Each
update of a particle consists of an attempt to move the physical
coordinate $\mathbf{r}_i$ together with the first one quarter of the path
variables $\mathbf{\bar{a}}_i$ (that is, together with the variables
$\left\{\mathbf{a}_{i,k}; 1 \leq k \leq [n_v/4]\right\}$) followed by a
separate attempt to move the rest of the path variables associated with
the particle $i$. Both the physical coordinates and the path variables
are moved in a cube centered about the old coordinates:
\[
\mathbf{r}'_i = \mathbf{r}_i + \Delta_r (2 \mathbf{u} - 1)
\]  and
\[
\mathbf{a}'_{i,k} =\mathbf{a}_{i,k} + \Delta_a(2 \mathbf{u} -1),
\] where the three components of $\mathbf{u}$ are independent uniformly
distributed random numbers on the interval $[0,1]$. Throughout our
simulations, we have used the following maximum displacement values:
$\Delta_r = 0.26$ \AA \, and $\Delta_a = 0.15$. The sampling technique
employed guaranties an acceptance ratio between 30\% and 70\%  for all
methods studied and for $n_v \leq 256$.

Because the acceptance ratio drops below 20\% for the Wiener-Fourier
reweighted technique with $n_v = 512$, each most basic step of the
previously described algorithm has been decomposed into two successive
steps. The first step is decomposed into an attempt to move the physical
coordinate $\mathbf{r}_i$ together with the first $1/8$ of the path
variables $\mathbf{\bar{a}}_i$, followed by an attempt to move the
physical coordinate $\mathbf{r}_i$ together with the next $1/8$ path
variables $\mathbf{\bar{a}}_i$. The second step is decomposed in a
similar fashion; half of the remaining variables have been moved in a
first attempt and then the other half in a second attempt. This restores
the overall acceptance ratio to about 33\%. In fact, we have monitored
separately the acceptance ratio for the four different steps necessary to
update all the coordinates associated with a given particle and have made
sure that the sampling is well balanced in the sense that the acceptance
for each individual step is about 30\% or larger.

As a counting device, we define a \emph{pass} as the minimal set of Monte
Carlo attempts over all variables in the system. A pass consists of $2
\cdot 22 = 44$ basic steps for all simulations with $n_v \leq 256$. For
the Wiener-Fourier reweighted technique with $n_v = 512$, a pass consists
of $4 \cdot 22 = 88$ basic attempted moves. One also defines a
\emph{block} as a computational unit made up of ten thousand passes.

\subsection{Embarrassingly parallel computation}

In order to achieve a statistical error of about $0.1$ K/molecule for all
computed energies,  we have employed a large number of Monte Carlo passes
(10.4 million) and we have divided the computation in $16$ independent
tasks  to be run in parallel. For the Wiener-Fourier reweighted method with $n_v = 512$, we have utilized a number of 40 million passes divided in $80$ independent tasks. The Monte Carlo simulations are embarrassingly
parallel provided that one can generate independent streams of uniformly
distributed random numbers. In this situation, there is no need for
communication among the different code replica running on different
nodes, and the program is an ideal candidate for use on a distributed computing
cluster. However, to be mathematically rigorous, it is necessary to
ensure that all the communication needed is already buried in the
independence of the streams of random numbers. This underlies the need
for ``good'' parallel random number generators.

The Mersenne Twister (MT) is a fast serial pseudorandom number generating
algorithm with a long period and good $k$-distribution
properties.\cite{Mat98} Quite interestingly, the algorithm allows for the
development of random number generators meeting certain user
specifications. For instance, one may specify the period (which must be a
Mersenne prime number i.e., a prime number of the form $2^p-1$), the word
size, or the memory size. Given a specified period, one may still produce
various algorithms which differ by their characteristic polynomials. The
dynamic creation of distributed random number generators is based on the
hypothesis that MT random number generators having relatively prime
characteristic polynomials produce highly independent streams of random
numbers.\cite{Mat98a} Because the laws by which the numbers are generated
are significantly different, it is very probable that the streams
produced by the different generators are highly uncorrelated. In this
paper, we have used the Dynamic Creator C-language library\cite{Mat98b}
with the Mersenne number $2^{3217}-1$. The library outputs streams of
32-bit integers, which are easy to convert into real numbers on the
interval $[0,1]$. Different streams are identified by different
identification numbers. The streams have been initialized once at the
beginning of the simulation with different seeds.

Given the $16$ streams of independent random numbers, the Monte Carlo
simulation proceeds as follows. For each stream, one performs an
independent simulation consisting of $65$ blocks. These blocks are
preceded by $13$ equilibration blocks, which are needed to bring the
system into probable configurations but do not contribute to the averages
of the estimating functions. For  the Wiener-Fourier reweighted method
with $n_v = 512$, we use $80$ independent streams of $50$ blocks each,
for a total of $40$ million passes. The equilibration phase consists of
$10$ blocks for each stream.  Ideally, the length of the individual
streams should be chosen to be sufficiently large, that the averages of
the computed property for different streams are independent and normally
distributed, as dictated by the central limit theorem. This requirement
is satisfied by all simulations we have performed.

We have collected individual averages for all blocks and streams  and
performed several statistical tests verifying the applicability of the
central limit theorem  as well as the independence between the block
averages of same or different streams. Let $\{Z_{i,j}: 1\leq i \leq 16;
1\leq j \leq 65\}$ denote the block-averages of the property $Z$ for
stream $i$ and block $j$ (the RW-WFPI simulation for $n_v = 512$ has been
analyzed in a similar fashion). Under the assumption that the size of the
blocks is large enough so that the correlation between different
block-averages is negligible and under the assumption that the
block-averages for different streams are highly uncorrelated, the values
$Z_{i,j}$ should have a Gaussian distribution centered around the average
value
\begin{equation}
\label{eq:3.2}
\overline{Z} =\frac{1}{16 \cdot 65} \sum_{i=1}^{16} \sum_{j=1}^{65}
Z_{i,j}
\end{equation} with variance
\begin{equation}
\label{eq:3.3}
\sigma^2(Z) = \frac{1}{16 \cdot 65} \left(\sum_{i=1}^{16} \sum_{j=1}^{65}
Z_{i,j}^2\right) - \overline{Z}^2.
\end{equation} The validity of this assumption can be verified with the
help of the Shapiro-Wilks normality test.\cite{Sha65} If the collection
of samples
$Z_{i,j}$ does not pass the test, it does not necessarily follow that the
samples
$Z_{i,j}$ are not independent, as their distribution is normal only if
the size of the blocks is sufficiently large. At a significance level of
5\%,  we do not reject the Gaussian distribution hypothesis for all
computed average properties. To within the statistical significance of
our calculations, the samples $Z_{i,j}$ can be assumed to be independent
and have a Gaussian distribution.

A second set of tests consists in verifying that the row and column
averages of $Z_{i,j}$ have Gaussian distributions centered around
$\overline{Z}$ with variances $\sigma^2(Z)/65$ and $\sigma^2(Z)/16$,
respectively. The validity of this distribution follows from the central
limit theorem and the assumption that the samples $Z_{i,j}$ are
independent and have a Gaussian distribution characterized by the average
$\overline{Z}$ and the variance $\sigma^2(Z)$. It is important to
emphasize that the row averages must pass this test. As previously
discussed, the number of blocks in a stream should be sufficiently large
so that the row averages have the required distribution even if the
independent samples $Z_{i,j}$ do not have a Gaussian distribution. Again,
under the assumption of independence only, the row averages should have a
Gaussian distribution centered around $\overline{Z}$ and have variance
$\sigma^2(Z)/N_\text{blocks}$ for a sufficiently large number of blocks
$N_\text{blocks}$. We have employed the Kolmogorov-Smirnov
test\cite{Pre92} to compare the distributions of the row and column
averages with the theoretical Gaussian distributions. For all computed
average properties, we find that the respective distributions are
identical at a statistical significance level of 5\%. The agreement for
the distribution of the row averages is evidence that the streams
generated by the Dynamic Creator package are sufficiently independent,
whereas the agreement for the distribution of the column averages is
evidence that the block averages of the same streams are independent.

For the third set of tests, we have considered two time-series $\{Z'_i, 1
\leq i \leq 16 \cdot 65\}$ and $\{Z''_i, 1 \leq i \leq 16 \cdot 65\}$
obtained by concatenating the rows of the matrix $Z_{i,j}$ and the
columns, respectively. We then have studied the autocorrelation of the
two time series for a maximum lag of $32$. The correlation coefficients
for a lag $k \leq 32$ are computed with the formula
\[ r'_k = \frac{1}{\sigma^2(Z)} \frac{1}{16\cdot 65} \sum_{i = 1}^{16
\cdot 65} \left(Z'_i - \overline{Z}\right)
\left(Z'_{i+k}-\overline{Z}\right),
\]  where $Z'_{i + k} = Z'_{i + k - 16 \cdot 65}$ if $i + k > 16 \cdot
65$. Under the independence hypothesis of the samples $Z'_i$, the
statistics of the correlation coefficients for $1 \leq k \leq 32$ is
normal with average zero and standard deviation $\sigma' = 1/\sqrt{16
\cdot 65}$. Moreover, the correlation coefficients can be regarded as
independent samples of this normal distribution. By the binomial
distribution, the probability that at most $m$ correlation coefficients
lie outside the interval $[-2\sigma', 2\sigma']$ is given  by the formula
\[ P(m) = \sum_{k = 0}^{m}  \frac{32!}{k!(32 - k)!} q^k (1-q)^{32 -k },
\] where $q \approx 0.046$ is the probability that a normal distributed
variable of mean zero and standard deviation $\sigma'$ lies outside the
interval $[-2\sigma', 2\sigma']$. One computes $P(3) = 0.942$ and $P(4) =
0.985$ so at a level of significance of 5\%, the hypothesis that the
$r'_k$ are independent samples of a normally distributed variable of mean
zero and standard deviation $\sigma' = 1/\sqrt{16 \cdot 65}$ should be
rejected if  $4$ or more correlation coefficients lying outside the
interval $[-2\sigma', 2\sigma']$ are observed.

\begin{figure}[!tbp]
     \includegraphics[angle=270,width=8.5cm,clip=t]{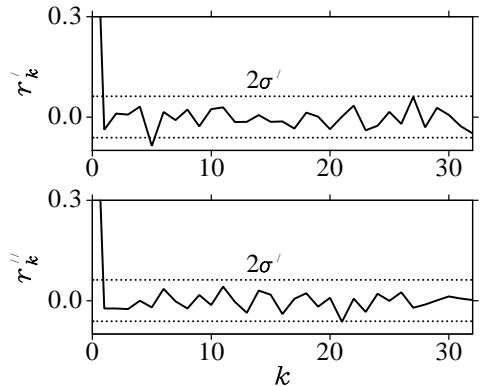}
   \caption[sqr] {\label{Fig:1} Correlograms for the time-series $Z'_i$ and
$Z''_i$. The property $Z$ is the  average ensemble energy computed by
means of the H-method estimator using the RW-WFPI method with $n_v = 32$.
One notices that both the correlation between the block averages ($r'_k$)
and the correlation between the streams ($r''_k$) are negligible. }
\end{figure}

The autocorrelation of the series $Z'_i$ is representative of the
correlation between the block averages of  same streams, whereas the
autocorrelation of the time series $Z''_i$ is representative of the
correlation between the blocks of similar rank corresponding to different
streams. Fig.~\ref{Fig:1} shows the correlograms of the two series for a
RW-WFPI Monte Carlo simulation with $n_v = 32$. The computed property is
the H-method energy estimator. Both series $Z'_i$ and $Z''_i$ have only
one point lying outside the interval $[-2\sigma', 2\sigma']$. These
points are $r'_5$ and $r''_{21}$, respectively (of course, the points
$r'_0 = r''_0 =1$ are not counted). Consequently, the simulation passes
our third statistical test. In fact, all the simulations performed have
passed this statistical test for all computed properties. We conclude
that the correlation between the block averages of same or different
streams is negligible. By the central limit theorem, the statistical
error in the determination of the average of the property  $Z$ is
\begin{equation}
\label{eq:3.4}
\pm 2\sigma(Z) / \sqrt{16 \cdot 65},
\end{equation} where $\sigma^2(Z)$ is defined by Eq.~(\ref{eq:3.3}). (For
the statistical error, we employ the $2 \sigma$ value, corresponding to
an interval of 95\% confidence. The 5\% probability that the results lie
outside the confidence interval is chosen to agree with the level of
significance of the statistical tests).

The analysis performed in the present subsection demonstrates that the
streams generated by the Dynamic Creator algorithm have negligible
correlation at least for our purposes.

A separate advantage in the use of independent streams is to overcome
the phenomenon of quasiergodicity,\cite{Val77} which might appear in
Monte Carlo simulations whenever the distribution that is sampled has
several well defined minima that are separated by walls of high energy.
In this case, the random walker may be trapped in one of the wells and
never sample the others, or sample them with the wrong frequency. The
Monte Carlo simulation may pass all the aforementioned statistical tests
but still produce the wrong results. For our system, the probability that
such a situation may occur is quite low because the system is highly
quantum mechanical with strong barrier tunneling. Moreover, the $16$
independent streams have been initialized randomly in configuration
space. This makes it unlikely that all the streams are trapped precisely
into the same local minimum or group of local minima. Evidence for
quasiergodicity may be captured in the form of a few outlying averages
among the stream averages. Such outlying averages have not been observed.

\subsection{Summary and discussion of the computed averages}
\begin{table*}[!bh]
\caption{Listed are the results obtained by the  Wiener-Fourier reweighted
path integral method. Average potential $\langle V \rangle_\beta$,
kinetic $\langle K \rangle_\beta$, and total energies $\langle E
\rangle_\beta$ are calculated with the help of the T- and H-estimators as
functions of the number of path variables $n_v$. The error bars are two
standard deviation values. All energies are given in K/molecule.}
\label{table:rw-wfpi}
\begin{tabular}{ccccccc}
   \hline \hline
$n_v$ \hs & $\langle E \rangle^T_\beta$ \hs & $\langle E
\rangle^H_\beta$
    \hs & $\langle V \rangle^T_\beta$ \hs & $\langle V \rangle^H_\beta$
    \hs & $\langle K \rangle^T_\beta$ \hs & $\langle K \rangle^H_\beta$
\\ \hline \\ 4  \hs& -57.66~$\pm$~0.05 \hs& -16.63~$\pm$~0.18 \hs&
-82.14~$\pm$~0.07
     \hs& -61.72~$\pm$~0.12 \hs&  24.48~$\pm$~0.02 \hs&  45.09~$\pm$~0.15
\\ \\ 8  \hs& -37.61~$\pm$~0.05 \hs& -17.77~$\pm$~0.16 \hs&
-64.74~$\pm$~0.06
     \hs& -53.07~$\pm$~0.11 \hs&  27.13~$\pm$~0.02 \hs&  35.29~$\pm$~0.13
\\ \\ 16 \hs& -25.68~$\pm$~0.04 \hs& -18.28~$\pm$~0.13 \hs&
-54.27~$\pm$~0.06
     \hs& -49.33~$\pm$~0.10 \hs&  28.60~$\pm$~0.03 \hs&  31.06~$\pm$~0.11
\\ \\ 32 \hs& -20.23~$\pm$~0.04 \hs& -18.00~$\pm$~0.12 \hs&
-49.66~$\pm$~0.06
     \hs& -48.05~$\pm$~0.10 \hs&  29.42~$\pm$~0.03 \hs&  30.05~$\pm$~0.11
\\ \\ 64 \hs& -18.29~$\pm$~0.04 \hs& -17.85~$\pm$~0.11 \hs&
-48.19~$\pm$~0.06
     \hs& -47.86~$\pm$~0.09 \hs&  29.90~$\pm$~0.03 \hs&  30.01~$\pm$~0.11
\\ \\ 128\hs& -17.75~$\pm$~0.04 \hs& -17.64~$\pm$~0.12 \hs&
-47.83~$\pm$~0.06
     \hs& -47.81~$\pm$~0.09 \hs&  30.08~$\pm$~0.03 \hs&  30.17~$\pm$~0.11
\\ \\ 256\hs& -17.71~$\pm$~0.04 \hs& -17.70~$\pm$~0.12 \hs&
-47.85~$\pm$~0.07
     \hs& -47.87~$\pm$~0.10 \hs&  30.14~$\pm$~0.03 \hs&  30.17~$\pm$~0.12
\\ \\
     \hline \hline
\end{tabular}
\end{table*}

\begin{table*}[!bh]
\caption{Listed are the results obtained by the  L\'evy-Ciesielski
reweighted path integral method. The format is that of Table
\protect\ref{table:rw-wfpi}.}

\label{table:rw-lcpi}
\begin{tabular}{ccccccc}
   \hline \hline
$n_v$ \hs & $\langle E \rangle^T_\beta$ \hs & $\langle E
\rangle^H_\beta$
    \hs & $\langle V \rangle^T_\beta$ \hs & $\langle V \rangle^H_\beta$
    \hs & $\langle K \rangle^T_\beta$ \hs & $\langle K \rangle^H_\beta$
   \\ \hline \\ 3  \hs& -70.46~$\pm$~0.06 \hs&  18.24~$\pm$~0.20 \hs&
-93.47~$\pm$~0.07
     \hs& -69.03~$\pm$~0.09 \hs&  23.01~$\pm$~0.02 \hs&  87.27~$\pm$~0.19
\\ \\ 7  \hs& -44.08~$\pm$~0.05 \hs& -10.81~$\pm$~0.15 \hs&
-71.03~$\pm$~0.06
     \hs& -55.08~$\pm$~0.08 \hs&  26.94~$\pm$~0.02 \hs&  44.28~$\pm$~0.14
\\ \\ 15 \hs& -29.84~$\pm$~0.04 \hs& -15.84~$\pm$~0.12 \hs&
-58.33~$\pm$~0.06
     \hs& -49.10~$\pm$~0.07 \hs&  28.50~$\pm$~0.02 \hs&  33.26~$\pm$~0.12
\\ \\ 31 \hs& -22.76~$\pm$~0.04 \hs& -17.40~$\pm$~0.10 \hs&
-51.95~$\pm$~0.06
     \hs& -47.83~$\pm$~0.06 \hs&  29.19~$\pm$~0.03 \hs&  30.43~$\pm$~0.11
\\ \\ 63 \hs& -19.50~$\pm$~0.04 \hs& -17.68~$\pm$~0.10 \hs&
-49.15~$\pm$~0.06
     \hs& -47.69~$\pm$~0.06 \hs&  29.65~$\pm$~0.03 \hs&  30.01~$\pm$~0.11
\\ \\ 127\hs& -18.25~$\pm$~0.04 \hs& -17.68~$\pm$~0.10 \hs&
-48.20~$\pm$~0.06
     \hs& -47.80~$\pm$~0.06 \hs&  29.95~$\pm$~0.03 \hs&  30.11~$\pm$~0.11
\\ \\ 255\hs& -17.84~$\pm$~0.04 \hs& -17.65~$\pm$~0.11 \hs&
-47.93~$\pm$~0.07
     \hs& -47.85~$\pm$~0.07 \hs&  30.09~$\pm$~0.03 \hs&  30.20~$\pm$~0.12
\\ \\
     \hline \hline
\end{tabular}
\end{table*}

\begin{table*}[!bh]
\caption{Listed are the results obtained by the trapezoidal Trotter
discrete path integral method. The format is that of Table
\protect\ref{table:rw-wfpi}.}

\label{table:lc-ttpi}
\begin{tabular}{ccccccc}
   \hline \hline
$n_v$ \hs & $\langle E \rangle^T_\beta$ \hs & $\langle E
\rangle^H_\beta$
    \hs & $\langle V \rangle^T_\beta$ \hs & $\langle V \rangle^H_\beta$
    \hs & $\langle K \rangle^T_\beta$ \hs & $\langle K \rangle^H_\beta$
    \\ \hline \\ 3  \hs& -68.54~$\pm$~0.05 \hs&  78.08~$\pm$~0.30 \hs&
-89.88~$\pm$~0.07
     \hs& -89.88~$\pm$~0.07 \hs&  21.34~$\pm$~0.02 \hs& 167.97~$\pm$~0.32
\\ \\ 7  \hs& -45.29~$\pm$~0.05 \hs&   7.22~$\pm$~0.19 \hs&
-70.88~$\pm$~0.06
     \hs& -70.88~$\pm$~0.06 \hs&  25.58~$\pm$~0.02 \hs&  78.10~$\pm$~0.21
\\ \\ 15 \hs& -30.61~$\pm$~0.04 \hs& -12.52~$\pm$~0.13 \hs&
-58.53~$\pm$~0.06
     \hs& -58.53~$\pm$~0.06 \hs&  27.92~$\pm$~0.02 \hs&   46.01~$\pm$~0.15
\\ \\ 31 \hs& -22.95~$\pm$~0.04 \hs& -16.86~$\pm$~0.11 \hs&
-51.99~$\pm$~0.06
     \hs& -51.99~$\pm$~0.06 \hs&  29.04~$\pm$~0.03 \hs&   35.14~$\pm$~0.12
\\ \\ 63 \hs& -19.55~$\pm$~0.04 \hs& -17.66~$\pm$~0.10 \hs&
-49.19~$\pm$~0.06
     \hs& -49.19~$\pm$~0.06 \hs&  29.65~$\pm$~0.03 \hs&   31.53~$\pm$~0.11
\\ \\ 127\hs& -18.29~$\pm$~0.04 \hs& -17.70~$\pm$~0.10 \hs&
-48.27~$\pm$~0.06
     \hs& -48.27~$\pm$~0.06 \hs&  29.97~$\pm$~0.03 \hs&   30.57~$\pm$~0.11
\\ \\ 255\hs& -17.86~$\pm$~0.04 \hs& -17.71~$\pm$~0.11 \hs&
-47.94~$\pm$~0.07
     \hs& -47.94~$\pm$~0.07 \hs&  30.07~$\pm$~0.03 \hs&   30.23~$\pm$~0.12
\\ \\
     \hline \hline
\end{tabular}
\end{table*}

The computed averages for all methods and estimators utilized are
presented in  Tables~\ref{table:rw-wfpi}, \ref{table:rw-lcpi}, and
\ref{table:lc-ttpi}.  For a given number of path variables
$n_v$, the RW-WFPI, RW-LCPI, and TT-DPI methods utilize $2n_v$,
$2.25 n_v$, and $n_v$ quadrature points, respectively. [For a discussion
of the minimal number of quadrature points and of the nature of the
quadrature schemes that must be employed for the first two methods, the
reader should consult Ref.~\onlinecite{Pre03b}. For the RW-WFPI method,
we have utilized $2n_v$ Gauss-Legendre quadrature points, though a number
of
$1.75n_v$ points would have sufficed.] The observed overall computational
time for the three methods have followed the ratios $2 : 2.25 : 1$, even
though the time necessary to compute the paths is proportional to $n_v^2$
for the first method and to $n_v \log_2(n_v)$ for the other methods. The
computation of the paths takes full advantage of the vector floating
point units of the modern processors and is dominated by the calls to the
potential, except for very large $n_v$.

As discussed in Ref.~\onlinecite{Pre03b}, the asymptotic convergence
for the reweighted techniques is expected to be cubic, even for the
Lennard-Jones potential that is not included in the class of potentials
for which cubic convergence has been demonstrated formally. We find that
the asymptotic convergence is attained only for very large $n_v$, as one may see by comparing
for example the total, potential, and kinetic energies computed with the
help of the T-method estimator for the RW-LCPI and the TT-DPI methods.
Even if the latter method has only $1/n_v^2$ asymptotic convergence, the
two methods produce almost equal results. In fact, a numerical analysis
of the relationship
\[
\left\langle E \right\rangle_{\beta,n_v}^T \approx \left\langle
E\right\rangle_\beta + \frac{const}{(n_v)^\alpha},
\] in which the left-hand side quantity is plotted against
$1/(n_v)^\alpha$ for different values of $\alpha$,  suggests that, while
the methods have converged within the statistical error, none of the
three methods includes sufficiently large values of $n_v$  to attain the
ultimate asymptotic rate of convergence.

When comparing the values of the H-method energy estimator and of the
related potential and kinetic estimators for the three path integral
techniques, one notices that the RW-LCPI technique provides  better
values than the TT-DPI method. The H-method estimator has a better behavior if used together with a reweighted technique.  This behavior is
consistent with the analysis we have performed in Section~II on the
values of the potential point-estimators and the excellent values found
with the RW-WFPI method. For the reweighted techniques, the H-method
estimator provides better energy values than the T-method estimator. This
is also true of the potential and kinetic parts of the estimators.
However, the variance of the H-method estimator is significantly larger
than the variance of the T-method estimator and the difference is even
more pronounced if one compares the corresponding kinetic estimators.

As discussed in Section~II.A, the path estimator for the potential energy has a smaller variance than the point estimator. Indeed, the results from Table~\ref{table:rw-wfpi} show that the variance of the path estimator is approximately $(0.9/0.6)^2=2.25$ times smaller than the variance of the point estimator. In the case of the RW-LCPI and TT-DPI methods, we have employed the estimator given by Eq.~(\ref{eq:24}) and the path estimator, respectively. These were shown to produce values identical to the point estimator but  have the variance of the path estimator. For the RW-WFPI and RW-LCPI methods, the point and the path estimators produce different results. Due to the very design of the reweighted techniques, we have argued that the point estimator results should be the more accurate ones. This theoretical prediction is well supported by the values presented in Tables~\ref{table:rw-wfpi} and~\ref{table:rw-lcpi}.

While we have argued that the H-method estimator is a better estimator as value (but not necessarily as variance) than the T-method estimator for the reweighted methods, it is apparent from Table~\ref{table:lc-ttpi} that the same difference persists for the trapezoidal Trotter scheme. As discussed before, for the TT-DPI method, the point and path estimators provide the same value for the average potential. As opposed to the reweigthed techniques, the H-method kinetic estimator is less accurate than the T-method kinetic energy estimator. Quite interestingly, even if individually the potential and the kinetic parts are more accurate for the T-method estimator, it is the H-method energy estimator that provides a more accurate total energy. Clearly, a strong compensation of errors appears in the case of the H-method estimator. Such a compensation of errors is generally characteristic of variational methods. In this respect, notice that the TT-DPI density matrices are positive definite because they are obtained by Lie-Trotter composing a certain symmetrical short-time approximation. By the Ritz variational principle, the H-method energy estimator cannot have a value smaller than the ground-state energy. Thus, the Ritz variational principle provides some control on the values of the H-method estimator, but not on the individual components, nor on the T-method estimator. The RW-LCPI density matrices are also positive definite for $n \geq 2$ and indeed, the energy provided by the H-method estimator is still better than what the values of the potential and kinetic parts suggest. While a final resolution awaits further study, it is apparent that this finding is not related to the asymptotic rate of convergence of the path integral technique. 

Among the three methods presented, the RW-WFPI has the fastest
convergence for all properties studied. Moreover, for $n_v = 128$ and
$n_v = 256$, there is a good agreement (within statistical noise) between
the T- and the H-method energy estimators, as well as between their
potential and kinetic energy components. For $n_v = 256$, one concludes
that the systematic error is  smaller than the statistical error for all
properties computed. An additional RW-WFPI simulation with $n_v = 512$
in $40$ million Monte Carlo passes has produced results consistent with the findings above. The results are
summarized in Table~\ref{tab:1} and represent the energy values we report.
\begin{table}[!htbp]
\caption{Estimated energies in K/molecule for the $(\text{H}_2)_{22}$
cluster computed with the help of the Wiener-Fourier reweighted technique
using $512$ path variables and $40$ million Monte Carlo passes. Listed are the average potential $\langle V \rangle_\beta$,
kinetic $\langle K \rangle_\beta$, and total energies $\langle E
\rangle_\beta$ calculated with the help of the T-method (left column) and
H-method (right column) estimators. The reported errors are two standard
deviations.}
\label{tab:1}
\begin{tabular}{c  r c l | c  r c l }
\hline  \hline
$\langle E \rangle^T_\beta \ $  &  $-17.69$ & $\pm$ & $ 0.02 \ $  & $ \
\langle E \rangle^H_\beta \ $  & $-17.71 $ & $\pm$ & $  0.06$
\vspace{0.75mm} \\
$\langle V \rangle^T_\beta \ $  &  $-47.82$ & $\pm$ & $ 0.03$ & $ \
\langle V \rangle^H_\beta \ $  & $-47.81 $ & $\pm$ & $  0.05$
\vspace{0.75mm} \\
$\langle K \rangle^T_\beta \ $  &  $30.13$ & $\pm$  & $0.02$ & $ \
\langle K \rangle^H_\beta \ $  & $30.10 $ & $\pm$ & $  0.06$  \\
\hline \hline
\end{tabular}
\end{table}

\section{Conclusions} \label{sec:conclude}

In the present work we have considered a number of issues related to the
choice of estimators for random series path integral methods.  We have
illustrated our results by applying them to the problem of computing
various thermodynamic properties of a model of the (H$_2)_{22}$  cluster
using reweighted path integral techniques.  The molecular hydrogen cluster
is a strongly quantum mechanical system and is representative of the type
of problems one is likely to encounter in many applications. Hence, it
constitutes a useful  benchmark for present and future path integral
techniques and for this reason it is important that its physical
properties be determined within advertized statistical error bars. Path
integral methods capable of dealing with such highly quantum-mechanical
systems in an efficient manner are needed, both for reliable
determinations of the physical properties of the respective systems as
well as for accurate parameterizations of the intermolecular potentials.

We wish to make a number of points concerning the present results and the
methods we have utilized to obtain them.  At a more general level, we
would like to emphasize that the reweighted path integral methods
discussed here provide a broadly applicable, simple, and formally well
characterized set of techniques.  As demonstrated by the present results,
they are capable of producing high-quality numerical results for problems
of appreciable physical complexity.  Moreover, they do so without the
assumption of a particular form for the underlying microscopic forces.
Furthermore, the estimators described in the present paper are convenient,
accurate, and easily implemented for any random series approach.  As
discussed in Section III, when used together, the T and H-method
estimators provide an important consistency check on the quality of the
path integral simulations.  Such consistency checks are a valuable
element in judging the reliability of particular simulations.

As previously mentioned, the cluster application discussed here provides
a convenient test bed for the development of numerical methods.  For this
reason, we have exercised due diligence with respect to the quality of
our final results summarized in Table~\ref{tab:1}.  As discussed in Section III, we
have subjected both the parallel random number generator employed and the
numerical results obtained to a series of quality-control tests.  Beyond
these statistical checks, it is important to note there is an internal
consistency check on the quality of the present results.  Specifically,
as is evident in Tables~\ref{table:rw-wfpi}, \ref{table:rw-lcpi}, and
\ref{table:lc-ttpi}, the kinetic, potential, and total energies
from the three different path integral approaches (trapezoidal Trotter,
reweighted L\'evy-Ciesielski, and reweighted Wiener-Fourier) all agree.
It is also important to note in this context that, while the presently
computed total energies agree with those reported by Chakravarty \emph{et
al.},\cite{Cha98} the individual kinetic and potential energies do not.
The kinetic energy reported by Chakravarty
\emph{et al.}\cite{Cha98} is approximately 0.8 K/particle higher than
found in the present simulations (with the
potential energy being correspondingly lower).  The magnitude of
this difference is well outside the statistical error bars involved and
appears to signal a systematic error.  Based on the observed
consistency between the results produced by
three different path integral methods and on the agreement between
the T and H-method estimators for each of these path integral
formulations, we feel confident of the results we have reported in Table~\ref{tab:1}.

\emph{Note:} After the present simulations had been completed, we have learned from D. M. Ceperley that the off-diagonal pair density used as the starting point in the simulations reported in Ref.~\onlinecite{Cha98} was truncated at  first order  in the expansion of off-diagonal displacements instead of second order and that the inclusion of this second order term resolves the kinetic and potential energy difference noted above.

\begin{acknowledgments} The authors acknowledge support from the National
Science Foundation through awards No. CHE-0095053 and CHE-0131114. They
also wish to thank the Center for Advanced Scientific Computing and
Visualization (TCASCV) at Brown University, especially Dr. James O'Dell, for valuable assistance with
respect to the numerical simulations described in the present paper. 
They would also like to thank  Mr. Cristian Diaconu for helpful
discussions concerning the present work.   Finally, the authors would
like to express a special thanks to Professor David Ceperley for
continuing discussions concerning the present simulations and for
his efforts in tracking down the origin of the pair density issues noted
in Section IV. 
\end{acknowledgments}

\appendix

\section{} The main purpose of this section is to give a compact form for
the integral
\begin{eqnarray}
\label{eq:A1}&&
\int_{\Omega}\ud P[\bar{a}]X_\infty(x,x',\bar{a};\beta)O[x_r(\theta)+
\sigma B_\theta^0(\bar{a})],
\end{eqnarray} where $\theta$ is an arbitrary point in the interval
$[0,1]$. In terms of a standard Brownian motion [see Eq.~(\ref{eq:1a})],
the above integral can be put into the form
\begin{widetext}
\begin{eqnarray}
\label{eq:A2}&& \nonumber P \left[\sigma B_1 = x' \big| \sigma B_0 = x
\right]\mathbb{E} \left[e^{-\beta \int_0^1 V(\sigma B_u)\ud u }O(\sigma
B_\theta) \Big| \sigma B_1 = x' , \sigma B_0 = x  \right] \\ && =
\int_{\mathbb{R}} \ud y O(y) P \left[\sigma B_1 = x', \sigma B_\theta = y
\big| \sigma B_0 = x \right]\mathbb{E} \left[e^{-\beta \int_0^1 V(\sigma
B_u)\ud u } \Big| \sigma B_1 = x' , \sigma B_\theta = y, \sigma B_0 = x
\right].
\end{eqnarray}
\end{widetext} Using the Markov property of the Brownian motion, one
readily justifies the equalities
\begin{eqnarray}
\label{eq:A3}&& \nonumber P\left[\sigma B_1 = x', \sigma B_\theta = y
\big| \sigma B_0 = x \right]\\ &&  = P\left[\sigma B_1 = x'\big| \sigma
B_\theta = y \right] P\left[ \sigma B_\theta = y \big| \sigma B_0 = x
\right] \qquad \\ && \nonumber  =
\rho_{fp}(x,y, \theta \beta) \rho_{fp}[y,x'; (1-\theta)\beta]
\end{eqnarray} and
\begin{eqnarray}
\label{eq:A4}&& \nonumber
\mathbb{E} \left[e^{-\beta \int_0^1 V(\sigma B_u)\ud u } \Big| \sigma B_1
= x' , \sigma B_\theta = y, \sigma B_0 = x \right] \\ && =\mathbb{E}
\left[e^{-\beta \int_0^\theta V(\sigma B_u)\ud u } \Big| \sigma B_\theta
= y , \sigma B_0 = x\right] \\ && \times \mathbb{E} \left[e^{-\beta
\int_{\theta}^1 V(\sigma B_u)\ud u } \Big| \sigma B_1 = x' , \sigma
B_\theta = y\right].\nonumber
\end{eqnarray}

Performing the transformation of coordinates $u'= u -\theta$ in the
second factor of the right-hand side of Eq.~(\ref{eq:A4}) and employing
the invariance of the Brownian motion under time translation
\begin{eqnarray*}&&
\left\{\sigma B_{u+\theta}\big| \sigma B_\theta = y , \sigma B_1 = x', u
\geq 0\right\}  \\ && \stackrel{d}{=} \left\{\sigma B_{u}\big| \sigma B_0
= y , \sigma B_{1-\theta} = x', u\geq 0\right\},
\end{eqnarray*} one obtains
\begin{eqnarray}
\label{eq:A5}\nonumber
\mathbb{E} \left[e^{-\beta \int_0^1 V(\sigma B_u)\ud u } \Big| \sigma B_1
= x' , \sigma B_\theta = y, \sigma B_0 = x \right] \\ =\mathbb{E}
\left[e^{-\beta \int_0^\theta V(\sigma B_u)\ud u } \Big| \sigma B_\theta
= y , \sigma B_0 = x\right] \\ \times \mathbb{E} \left[e^{-\beta
\int_0^{1-\theta} V(\sigma B_u)\ud u } \Big| \sigma B_{1-\theta} = x' ,
\sigma B_0 = y\right].\nonumber
\end{eqnarray} Let us focus on the term
\[
\mathbb{E} \left[e^{-\beta \int_0^\theta V(\sigma B_u)\ud u } \Big|
\sigma B_\theta = y , \sigma B_0 = x\right].
\] Performing the substitution of variables $u' = u / \theta$ and
employing the scaling property of the Brownian motion
\begin{eqnarray*}&&
\left\{\sigma B_{u\theta}\big| \sigma B_0 = x , \sigma  B_\theta = y, u
\geq 0\right\}  \\ && \stackrel{d}{=} \left\{\sigma
\theta^{1/2}B_{u}\Big| \sigma \theta^{1/2}B_0 = x , \sigma \theta^{1/2}
B_{1} = y, u\geq 0\right\},
\end{eqnarray*} one proves
\begin{eqnarray}
\label{eq:A6}\nonumber &&
\mathbb{E} \left[e^{-\beta \int_0^\theta V(\sigma B_u)\ud u } \Big|
\sigma B_\theta = y , \sigma B_0 = x\right]\\&& \nonumber  = \mathbb{E}
\left[e^{-\beta \theta \int_0^1 V(\sigma \theta^{1/2} B_u)\ud u } \Big|
\sigma \theta^{1/2} B_1 = y , \sigma \theta^{1/2} B_0 = x\right] \\&& =
\rho(x,y; \theta \beta)/\rho_{fp}(x,y; \theta \beta).
\end{eqnarray} In a similar fashion, one demonstrates that
\begin{eqnarray}
\label{eq:A7}\nonumber
\mathbb{E} \left[e^{-\beta \int_0^{1-\theta} V(\sigma B_u)\ud u } \Big|
\sigma B_{1-\theta} = x' , \sigma B_0 = y\right]\\ = \rho\left[y,x';
(1-\theta) \beta\right]\big/\rho_{fp}\left[y,x'; (1-\theta) \beta\right].
\end{eqnarray}

We now combine Eqs.~(\ref{eq:A1}), (\ref{eq:A2}), (\ref{eq:A3}),
(\ref{eq:A5}), (\ref{eq:A6}), and (\ref{eq:A7}) to obtain
\begin{eqnarray}
\label{eq:A8} \nonumber
\int_{\Omega}\ud P[\bar{a}]X_\infty(x,x',\bar{a};\beta)O[x_r(\theta)+
\sigma B_\theta^0(\bar{a})] \\ =
\int_{\mathbb{R}} \rho(x,y;\theta \beta) \rho[y,x'; (1-\theta)\beta] O(y)
\ud y.
\end{eqnarray} With the help of Eq.~(\ref{eq:A8}) and by cyclic
invariance,
\begin{eqnarray}
\label{eq:A9} &&\nonumber
\int_{\mathbb{R}}\ud x \int_{\Omega}\ud
P[\bar{a}]X_\infty(x,\bar{a};\beta)O[x+ \sigma B_\theta^0(\bar{a})] \\
&&\nonumber  =
\int_{\mathbb{R}}\ud x\int_{\mathbb{R}} \ud y \rho(x,y;\theta \beta)
\rho[y,x; (1-\theta)\beta] O(y)  \\&&   = \int_{\mathbb{R}} \ud y
\rho(y,y;\beta)  O(y)  \\ &&\nonumber = \int_{\mathbb{R}}\ud x
\int_{\Omega}\ud P[\bar{a}]X_\infty(x,\bar{a};\beta)O(x).
\end{eqnarray} Moreover, since the function $O(x)$ is arbitrary, the last
identity also implies that the random variables $x$ and $x+ \sigma
B_\theta^0(\bar{a})$ have identical distribution functions under the
probability measure
\[
\frac{X_\infty(x,\bar{a};\beta)\ud x \, \ud
P[\bar{a}]}{\int_{\mathbb{R}}\ud x \int_{\Omega}\ud
P[\bar{a}]X_\infty(x,\bar{a};\beta)}.
\] By setting $O(x)=1$ in Eq.~(\ref{eq:A8}), one obtains the well-known
product formula
\begin{eqnarray}
\label{eq:A10}\rho(x,x';\beta) =  \nonumber
\int_{\Omega}\ud P[\bar{a}]X_\infty(x,x',\bar{a};\beta)\\ =
\int_{\mathbb{R}} \rho(x,y;\theta \beta) \rho[y,x'; (1-\theta)\beta] \ud
y,
\end{eqnarray} which is seen to be a consequence of some basic properties
of the Brownian motion.




\end{document}